\documentclass[journal]{IEEEtran}
\usepackage{color}
\usepackage{graphicx}
\usepackage{cite}
\usepackage{subfig}
\usepackage{bm, amsmath, amssymb}
\usepackage{url}
\usepackage{algorithm}
\usepackage{algpseudocode}
\usepackage{multirow}
\usepackage[colorinlistoftodos]{todonotes}
\usepackage{booktabs}
\usepackage{enumitem}
\usepackage{soul}
\soulregister\cite{7}

\newcommand{\xb}{\bm x^{(b)}}
\newcommand{\xhat}{\hat{\bm x}}
\newcommand{\xhatb}{\hat{\bm x}^{(b)}}

\begin{document}
\title{End-to-End Learning of Transmitter and Receiver Filters in Bandwidth Limited Fiber Optic Communication Systems}

\author{Søren~Føns~Nielsen,  % ~\IEEEmembership{Member,~IEEE,}
        Francesco~Da~Ros, %~\IEEEmembership{Fellow,~OSA,}
        Mikkel~N.~Schmidt and Darko~Zibar% <-this % stops a space
    \thanks{This work was supported by VILLUM FONDEN with grants VI-POPCOM (VIL54486),
MARBLE (VIL40555), and YIP OPTIC-AI (VIL29334)}%
    \thanks{S. F. Nielsen and M. N. Schmidt are with the Department of Applied Mathematics and Computer Science, Technical University of Denmark mail: sfvn@dtu.dk.}% <-this % stops a space
    \thanks{F. D. Ros and D. Zibar are with the Department of Electrical and Photonics Engineering, Technical University of Denmark}% <-this % stops a space
    \thanks{Manuscript received XXXX; revised XXXX.}}
    

\markboth{Journal of \LaTeX\ Class Files,~Vol.~14, No.~8, August~2015}%
{Shell \MakeLowercase{\textit{et al.}}: Bare Demo of IEEEtran.cls for IEEE Journals}



\maketitle

\begin{abstract}
This paper investigates the application of end-to-end (E2E) learning for joint optimization of pulse-shaper and receiver filter to reduce intersymbol interference (ISI) in bandwidth-limited communication systems. We investigate this in two numerical simulation models: 1) an additive white Gaussian noise (AWGN) channel with bandwidth limitation and 2) an intensity modulated direct detection (IM/DD) link employing an electro-absorption modulator. For both simulation models, we  implement a wavelength division multiplexing (WDM) scheme to ensure that the learned filters adhere to the bandwidth constraints of the WDM channels. Our findings reveal that E2E learning greatly surpasses traditional single-sided transmitter pulse-shaper or receiver filter optimization methods, achieving significant performance gains in terms of symbol error rate with shorter filter lengths. These results suggest that E2E learning can decrease the complexity and enhance the performance of future high-speed optical communication systems.
\end{abstract}% add that we compare to Volterra (best-case, achievable...)


% \fi
\IEEEpeerreviewmaketitle

\section{Introduction}
% Digital age, datacenters, internet, power usage...
\IEEEPARstart{A}{s} we continue to integrate digital technology into all aspects of society, the traffic demands are estimated to grow at an exponential rate towards 2030~\cite{andraeNewPerspectivesInternet2020}. This in turn requires hyperscale datacenters (DC) located all around the world with high-speed optical interconnects, where intensity modulation with direct detection (IM/DD) is the most widely used technology. 

Future DC solutions will likely employ higher order PAM signaling and baud rates reaching 100 GBd or beyond to achieve 800 Gb/s and 1.6 Tb/s~\cite{berikaaNextgenerationObandCoherent2024}. This will require ultra-high bandwidths of optical and electrical components, which may be challenging to realize in practice. It is therefore expected that the next generation data center fiber-optic communication systems will need to deal with strong inter-symbol-interference (ISI), while at the same time keeping the complexity, and thereby power consumption, down~\cite{cheModulationFormatDigital2023}. The focus of the next generation of transmitter- and receiver-side digital signal processing (DSP) techniques should therefore focus on low complexity solutions for ISI mitigation.  

% Realizing zero-ISI
As we move towards 100 GBd or higher data rates, the bandwidth limitations of the transmitter and the receiver front-end components (digital-to-analog converter, optical modulator, photodiodes and analog-to-digital converter), along with chromatic dispersion in the fiber channel, will lead to strong ISI.

Realizing a zero ISI system is in practice a difficult problem, as it requires that the total transfer function---including pre-distortion, pulse-shaper, transmitter front-end, channel, receiver front-end, receiver (matched) filter and equalizer---must fulfill the Nyquist criterion~\cite{proakisDigitalCommunications2008}. Typically, the components that can be adjusted to meet the Nyquist criterion are the pre-distortion, the pulse-shaper, the receiver filter and the equalizer. On top of that, it is desirable to find a parsimonious solution promoting low complexity for many practical applications.

% Current solutions...
% Equalization...
Current solutions, for mitigating ISI are predominantly focused on receiver-side \emph{equalization}, i.e. the process of removing the distortion introduced by the system, which can be done using an adaptive linear filter, denoted a feed-forward equaliser (FFE)~\cite{proakisDigitalCommunications2008}. Feedback filter structures are often used in conjunction with the FFE, either after the FFE at the receiver, denoted decision feedback equalizer (DFE)~\cite{yuDecisionFeedbackEqualizerOptically2007a}, or at the transmitter, denoted Tomlinson-Harashima Precoding (THP)~\cite{rathTomlinsonHarashimaPrecoding2017}.
In systems with non-linear distortion due to hardware imperfections and fiber dispersion, non-linear compensation can be employed. Examples include non-linear adaptive filters such as the Volterra equalizer~\cite{stojanovicVolterraWienerEqualizers2017} and neural networks (see~\cite{karanovMachineLearningShort2022} for an overview).
% Pulse-shaping
Some effort has also gone into mitigating distortion from the transmitter side, either through \emph{pulse-shaping}~\cite{czeglediBandlimitedPowerEfficientSignaling2014}, i.e. designing spectrally efficient filters to minimize ISI, or predistortion~\cite{wuCband120GbPAM42022, bajajDeepNeuralNetworkBased2022} and constellation shaping~\cite{liangGeometricShapingDistortionLimited2023} to linearize the transmitter response. Typically these compensation mechanisms are designed or optimized independently of each other, which may lead to suboptimal performance. 

End-to-end learning (E2E)\cite{osheaIntroductionDeepLearning2017, karanovEndtoEndDeepLearning2018} has emerged as a framework for jointly optimizing all adaptable DSP blocks in the system on both the transmitter and receiver side. In \cite{karanovEndtoEndDeepLearning2018} this was shown for an IM/DD system, where both transmitter and receiver were modeled by a feed-forward neural network using an autoencoder-style structure. The system was trained offline (through numerical simulation) using the backpropagation algorithm given a known pilot sequence, and evaluated experimentally with a fine-tuning step for the receiver. This enabled information rates above 42 Gb/s beyond 40 km of transmission distance without amplification.
E2E has since been used in the context of geometric constellation shaping~\cite{jovanovicEndtoEndLearningConstellation2022}, joint geometric and probabilistic constellation shaping~\cite{neskorniukMemoryawareEndendLearning2023}, wavelength division multiplexing systems~\cite{songModelBasedEndtoEndLearning2022}, learning parameters in directly modulated laser transmission~\cite{hernandezEndtoendOptimizationOptical2024} and learning the pulse shape and receiver demapper in a 6G wireless application~\cite{marasingheWaveformLearningPhase2024}, to mention a few examples. 

However, to the best of our knowledge, no systematic analysis of E2E learning has been performed for the joint optimization of pulse-shaper and receiver filter, to reach the zero ISI criterion for AWGN and IM/DD channels under strong ISI. 

% In this paper,
In this paper we focus on the setting where the pulse-shaper and receiver filter are implemented as FIR filters, which is of great importance due to low complexity of implementation. We numerically investigate the use of E2E for jointly optimizing the pulse-shaper and receiver filter to combat ISI due to bandwidth limitations of the digital-to-analogue (DAC) and analogue-to-digital (ADC) converter, as well as chromatic dispersion, in a 100 GBd four pulse amplitude modulation (4-PAM) communication link. We first show the benefits of joint pulse-shaper and receiver filter optimization in a simple additive white Gaussian noise (AWGN) system with bandwidth limitation, where zero ISI can be reached for fairly short filter lengths. Next, we show the E2E framework's benefits in a wavelength division multiplexing (WDM) scheme, where the learned filters must adhere to the bandwidth restrictions imposed by the presence of interfering channels. We then show the benefits of joint optimization for a 1 km WDM IM/DD link with an electro-absorption modulator (EAM) that exhibits some degree of nonlinearity. In the non-linear case, we additionally compare our proposed joint optimization to a flexible Volterra equalizer with a large number of parameters to investigate the theoretical limits of this type of optimization. Finally, we evaluate the robustness of the learned filters to fiber non-linearities using the split-step Fourier method (SSFM).

% The rest of the paper...
The rest of the paper is structured as follows. In section~\ref{sec:methods}, we introduce the learning framework and the two communication links (AWGN and IM/DD) by their system model. Furthermore, we describe how WDM is implemented for both links. Next, in section~\ref{sec:results}, we present the results of our simulations, including symbol-error rate curves for different scenarios. Finally, in section~\ref{sec:conclusion}, we present our conclusions and perspectives on our findings.

\section{Numerical setup}\label{sec:methods}
We begin by introducing the learning framework which is based on stochastic gradient descent optimization. In the following sections, the employed end-to-end learning frameworks for AWGN and IM/DD are described.

\subsection{End-to-end Learning Using Stochastic Gradient Descent}\label{sec:preliminaries}
In the communication systems we investigate here, we are interested in obtaining pulse-shaper and receiver filters that jointly reduce ISI as much as possible. We employ a gradient based learning of the filter coefficients. Let the \emph{loss} function be denoted $\mathcal{L}(\bm x, \hat{\bm x}, \bm \theta)$, in which $\bm x$ is the true symbol sequence, $\xhat$ is the estimated symbol sequence before demapping and $\bm \theta = \left\{\bm h_\text{p}, \bm h_\text{r} \right\}$, is the collection of all parameters we want to optimize, in this case the pulse-shaper $\bm h_\text{p}$ and the receiver filter $\bm h_\text{r}$. In each iteration of the algorithm, we use gradient descent to minimize $\mathcal{L}$. This requires calculating the gradient of $\mathcal{L}(\cdot)$ with respect to $\bm \theta$, $\frac{\partial\mathcal{L}}{\partial\bm\theta}$. Furthermore, to avoid having to observe the entire sequence of symbols before updating the model, we use a stochastic gradient descent approach, processing only a batch of sybmols of size $N_\text{batch}$, denoted $\xb$. In all simulations, we use the mean-square error as our loss function, which for one batch, can be written as,
\begin{equation}
    \mathcal{L}(\hat{\bm x}^{(b)} , \bm x^{(b)}, \bm \theta) = \frac{1}{N_\text{batch}}\sum_{n=1}^{N_\text{batch}} (\hat{x}_n^{(b)} - x_n^{(b)})^2\label{eq:mse}
\end{equation}
Note here that the parameters $\bm \theta$ are implicitly used to produce $\hat{\bm x}$ but are not directly appearing in the right-hand side of \eqref{eq:mse}.

All the systems described below are implemented in PyTorch~\cite{paszkePyTorchImperativeStyle2019} which provides automatic differentiation for gradient computations using the backpropagation algorithm, as well as routines for stochastic gradient optimization. We use the Adam optimizer~\cite{kingmaAdamMethodStochastic2015}, which combines momentum and adaptive learning rates to achieve fast convergence and stable training. At each gradient update, we use gradient norm clipping~\cite{pascanuDifficultyTrainingRecurrent2013} and we normalize the filters to have unit L2-norm. We use a learning rate scheduling called OneCLR~\cite{smithSuperconvergenceVeryFast2019}. A high-level pseudo-code of our optimization routine can be seen in Alogrithm~\ref{alg:e2e-pseudo}, and our implementation is available online\footnote{Link to our Github repository:\\ \texttt{https://github.com/sfvnielsen/marble-end-to-end}}.

% End-to-end learning algortihm
\begin{algorithm}
    \caption{Optimization of FIR Filters}
    \label{alg:e2e-pseudo}
    \begin{algorithmic}[1]
    \State \textbf{Input:} Initial filters \(\bm h_\text{p}\), \(\bm h_\text{r}\), learning rate schedule \(\eta_b\), training data $\bm x$ and batch size $N_\text{batch}$.
    \For{$b = 1$ to $N_\text{symbols} / N_\text{batch}$}
        \State Evaluate system on $\xb$ to get predictions $\xhatb$.
        \State Calculate loss $\mathcal{L}_b = \mathcal{L}(\xb, \xhatb, \bm \theta)$.
        \State Compute the gradients: 
            $\bm g_\text{p} = \frac{\partial \mathcal{L}_b}{\partial \bm h_\text{p}}$, 
            $\bm g_\text{r} = \frac{\partial \mathcal{L}_b}{\partial \bm h_\text{r}}$,\newline
            \hspace*{3em}
            and apply norm clipping.
        \State Update parameters: \newline
        \hspace*{3em}
        $\bm h_\text{p} := \bm h_\text{p} - \eta_b \Delta_b(\bm g_\text{p})$, 
        $\bm h_\text{r} := \bm h_\text{r} - \eta_b \Delta_b(\bm g_\text{r})$,\newline
        \hspace*{3em}
        where $\Delta_b$ is the Adam~\cite{kingmaAdamMethodStochastic2015} update.
    \EndFor
    \State \textbf{Output:} Optimized filters $\bm h_\text{p}$, $\bm h_\text{r}$.
    \end{algorithmic}
\end{algorithm}    

\begin{figure*}
    \centering
    \subfloat[Training mode. Pulse-shaper and receiver filters are updated using backpropagation, visualized with the dashed red line.]{\label{fig:awgn-block-train}\includegraphics[width=\textwidth]{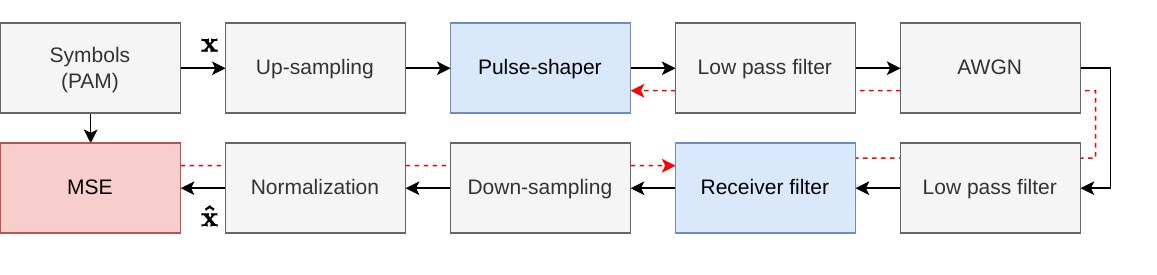}}\\
    \subfloat[Evaluation mode. Pulse-shaper and receiver filters are fixed to the values from last iteration of the training phase.]{\label{fig:awgn-block-eval}\includegraphics[width=\textwidth]{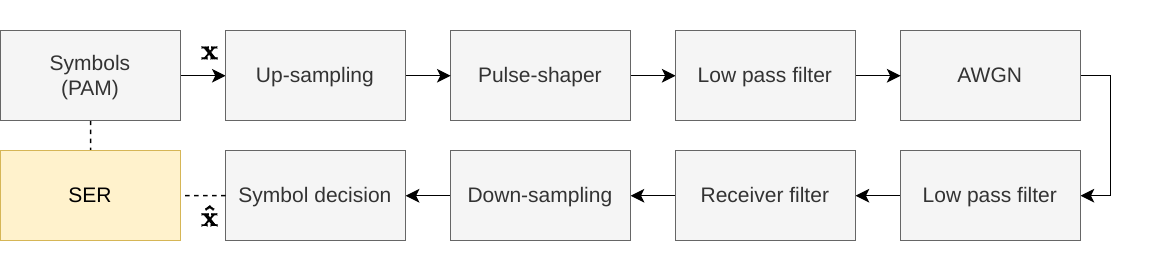}}
    \caption{Blockdiagram of the system model for the AWGN channel with bandwidth limitation. When filters are optimized (training) they are done so using the mean square error (MSE) between the output of the system and the symbol sequence, depicted in Figure~\ref{fig:awgn-block-train}. During evaluation (depicted in Figure~\ref{fig:awgn-block-eval}), a new symbol sequence is generated, filters are fixed and the symbol error rate (SER) is calculated.}
    \label{fig:awgn-block}
\end{figure*}

\subsection{Additive White Gaussian Noise Channel with Bandwidth Limitation}\label{sec:methods_awgn}
% AWGN with bandwidth-limitation
% Blockdiagrams
% Optimizers and costfunctions
% Training vs. eval
We first investigate end-to-end learning in a simple additive white Gaussian noise (AWGN) channel with bandwidth limitation, which is depicted in Figure~\ref{fig:awgn-block}. We assume that the bandwidth limitation originates from the DAC and ADC. The pulse-shaper and receiver filter are implemented as FIR filters with adjustable coefficients that are optimized using E2E learning. 

Pulse-amplitude modulation (PAM) symbols with constellation order 4 are drawn from a uniform distribution after which they are upsampled by a factor of 4. The up-sampled symbol sequence is passed through a finite impulse response (FIR) pulse-shaping filter such that at the output we have a symbol rate of $R_s = 100$ GBd.  To emulate the bandwidth limitation of the DAC, the signal is then low-pass filtered by a 5th order Bessel filter with 3dB cutoff $f_\text{3dB} = 45$ GHz. Next, AWGN is added such that the resulting signal's SNR can be varied. After the noise is added, another Bessel low-pass filter, emulating the bandwidth limitation of the ADC, is applied with same cutoff as the first one. A receiver FIR (matched) filter is applied and the signal is downsampled to a single sample per symbol (sps). Finally, we apply a normalization that rescales the signal to match the average power of the constellation. In all simulations, the pulse-shaper and receiver filters are initialized to a root-raised cosine (RRC) with a rolloff of $\rho = 0.01$ and we vary the length of the filters as part of the simulations. With the aforementioned initialization, it is apparent that due to the bandwidth limitations of the DAC and ADC, the total impulse response will not satisfy the Nyquist criterion. Therefore, the pulse-shaper and the matched filter coefficients need to be adapted.  

During training (see Figure~\ref{fig:awgn-block-train}), the output of the receiver filter, $\xhat$, is compared to the true symbol sequence $\bm x$ using MSE, as described in the previous section. The $\xhat$ and $\bm x$ are coarsely aligned before calculating the cost function, by a priori calculating the average group-delay of the two Bessel filters in the passband and compensating for that in the MSE calculation.

During evaluation (see Figure~\ref{fig:awgn-block-eval}), a new sequence of symbols is generated and propagated through the system with the learned filters fixed to their value from the last iteration of the training phase. The output of the downsampler is mapped to the closest constellation point in the Euclidean distance to obtain the symbol estimate, $\hat{\bm x}$. Finally the symbol error rate (SER) between the sequences $\bm x$ and $\hat{\bm x}$ is calculated. In the case of the AWGN, we train at one (high) SNR and evaluate over a range of SNRs to get performance curves. This can be done due to the linearity of the system (which is not true for the IM/DD system).

\subsection{Multilevel Intensity Modulated and Direct Detected System}\label{sec:methods_imdd}
% IM/DD from (Liang and Kahn)
Next, we investigate the end-to-end learning framework for an intensity modulated direct detection (IM/DD) link that employs an EAM and fiber transmission link.  We would like to investigate the benefits of joint pulse-shaper and receiver filter optimization in the presence of EAM nonlinearity and chromatic dispersion. Again, to keep the complexity low, the pulse-shaper and the receiver filter are implemented as FIR filters.

\begin{figure*}
    \centering
    \subfloat[Training mode. Pulse-shaper and receiver filters are updated using backpropagation, visualized with the dashed red line.]{\label{fig:imdd-block-train}\includegraphics[width=\textwidth]{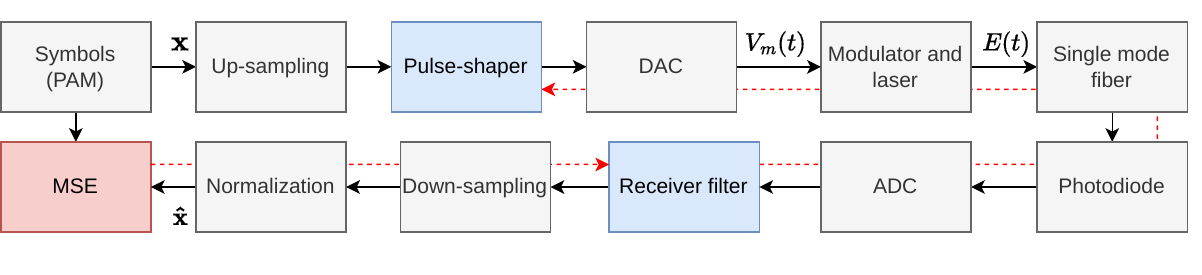}}\\
    \subfloat[Evaluation mode. Pulse-shaper and receiver filters are fixed to the values from last iteration of the training phase.]{\label{fig:imdd-block-eval}\includegraphics[width=\textwidth]{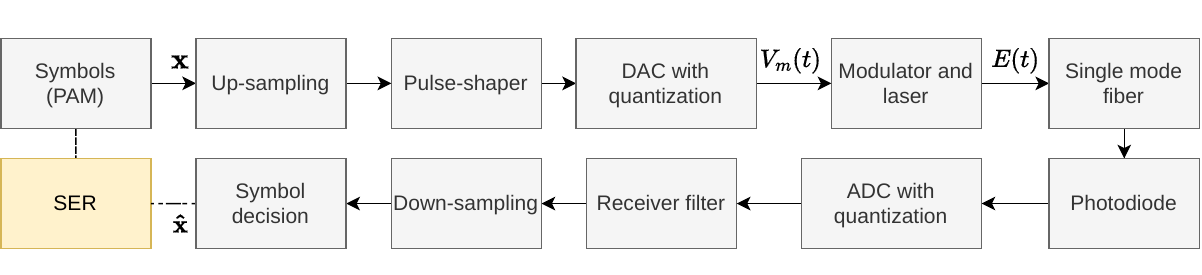}}
    \caption{System model for the simulated IM/DD link. When filters are optimized (training) they are done so using the mean square error (MSE) between the output of the system and the symbol sequence, depicted in Figure~\ref{fig:imdd-block-train}. During evaluation (depicted in Figure~\ref{fig:imdd-block-eval}), a new symbol sequence is generated, filters are fixed and the sybmol error rate (SER) is calculated. Additionally, quantization in the ADC and DAC is applied in the evaluation step.}
    \label{fig:imdd-block}
\end{figure*}

The block diagram for the system can be seen in Figure~\ref{fig:imdd-block}. We generate a PAM-4 symbol sequence, that after up-sampling at $sps = 8$ and pulse-shaping, has an $R_s = 100$ GBd. The DAC is comprised of a normalization step and a low-pass filter and outputs a voltage signal. We ignore quantization noise during training in the DAC. The normalization step is a multiplication with a gain, $g_\text{dac}$, after which a clipping function is applied with range $[-1/2, 1/2]$. A voltage peak-to-peak factor, $V_\text{pp}$, is now multiplied onto the signal such that it now has range $[-V_\text{pp}/2, V_\text{pp}/2]$. The low-pass filter is applied with same parameters as for the AWGN system (5th order Bessel, $f_\text{3dB} = 45$ GHz). 

The resulting voltage signal, $V(t)$, is combined with a bias voltage $V_\text{b}$ and applied to the modulator, i.e.~$V_\text{m}(t) = V_\text{b} + V(t)$.  We study two modulators, an ideal linear modulator and an electro-absorption modulator (EAM)\cite{liangGeometricShapingDistortionLimited2023, gallantCharacterizationDynamicAbsorption2008}.
The output of the ideal modulator has the form,
\begin{equation}
    E_\text{ideal}(t) = \sqrt{P_\text{in}^{(\text{ideal})} \cdot V_\text{m}(t)},\label{eq:ideal_mod}
\end{equation}
in which $P_\text{in}$ is the power of the laser and $E_\text{ideal}(t)$ is the resulting signal in the optical domain.

\begin{figure*}
    \centering
    \subfloat[Absorption curve $\alpha\left(V(t)\right)$. Red dots indicate the knee-points used to make the interpolation.]{\includegraphics[width=0.49\textwidth]{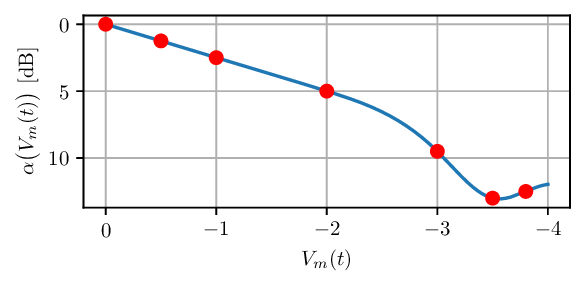}\label{fig:eam_absorption}}
    \subfloat[Relation between voltage and optical amplitude given by equation \eqref{eq:eam}.]{\includegraphics[width=0.49\textwidth]{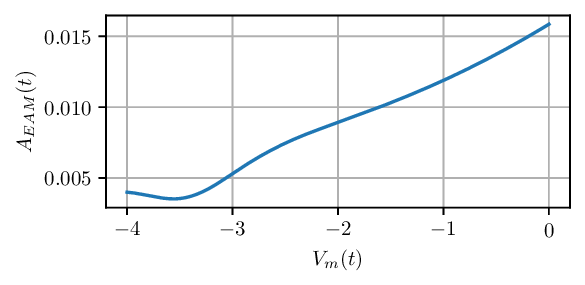}\label{fig:eam_transfer}}
    \caption{Electro-absorption modulator model used in the IM/DD system. Absorption curve was derived from \cite{liangGeometricShapingDistortionLimited2023} using a cubic spline fit.}
    \label{fig:eam}
\end{figure*}

In the case of the EAM, the amplitude of the signal at the output of the modulator is written as,
\begin{equation}
    A_\text{EAM}(t) = \sqrt{P_\text{in}^{(\text{EAM})} \cdot 10^{-\alpha\left(V_\text{m}(t)\right) / 10}},\label{eq:eam}
\end{equation}
in which $\alpha\left(V_\text{m}(t)\right)$ is the voltage-to-absorption function. We based $\alpha\left(V(t)\right)$ on the one reported in \cite{liangGeometricShapingDistortionLimited2023} using a cubic spline fit (cf. Figure~\ref{fig:eam_absorption}). To model the signal-intensity induced chirp, we follow the model from \cite{liangGeometricShapingDistortionLimited2023}, yielding following expression for the signal after the EAM,
\begin{equation}
    E_\text{EAM}(t) = A_\text{EAM}(t)\cdot e^{j \left(\frac{\alpha}{2} \ln A^2_\text{EAM}(t)\right)}, \label{eq:signal-after-eam}
\end{equation}
in which $\alpha$ is the linewidth enhancement factor that controls the amount of chirp. The voltage-to-optical amplitude relationship and the absorption curve for the EAM used in the simulations can be seen in Figure~\ref{fig:eam}.
For both types of modulators, we keep the laser power fixed at a given level (cf. Table~\ref{tab:imdd_system_parameters}) and only change $V_\text{pp}$ in our simulations.
% Single mode fiber
Chromatic dispersion and fiber attenuation, associated with the SMF, is modeled using a standard frequency domain filter~\cite{agrawalFiberopticCommunicationSystems2002}. We use the laser and fiber parameters from \cite{liangGeometricShapingDistortionLimited2023} (cf. Table~\ref{tab:imdd_system_parameters}).
% Photodiode
The signal is then detected using a square-law detection, with thermal noise and shot noise both modeled as white Gaussian random variables with variances $\sigma_\text{thermal}^2$ and $\sigma_\text{shot}^2$, respectively. The noise variances are parameterized as,
\begin{align}
    \sigma_\text{thermal}^2 &= \frac{4\cdot k \cdot T \cdot F_\text{s}}{B\cdot Z}, \\
    \sigma_\text{shot}^2 &= \frac{2 e_\text{c} \left( R\cdot \mathbb{E}[|y|^2] + I_\text{d}\right) F_\text{s}}{B},
\end{align}
in which $k$ is the Boltzmann constant, $T$ is the temperature of the photodiode in Kelvin, $F_\text{s}$ is the sampling frequency, $B$ is the bandwidth of the photodiode, $Z$ is the impedance load, $e_\text{c}$ is the electron charge, $R$ is the responsivity of the photodiode, $\mathbb{E}[|y|^2]$ is the empirical average power received and $I_\text{d}$ is the dark current.
% Rx
At the receiver, the signal is converted to digital domain using an ADC with the same low-pass filter characteristics as the DAC. The receiver filter is then applied and the signal is downsampled to $1$ sps and the resulting signal is normalized to match the average power of the constellation. An overview of the parameters and the values used in our experiments can be seen in Table~\ref{tab:imdd_system_parameters}.

The training and evaluation stages follow the same structure as in the previous section, namely, during training the MSE is calculated between the symbol sequence of interest and the output of the normalization block. The filters are updated using gradient descent as described in Algorithm~\ref{alg:e2e-pseudo}. During evaluation, we add a quantization step, with $2^5$ levels, to the DAC and ADC. The downsampled output of the receiver filter is mapped to its closest symbol in the constellation and the SER is calculated.

% Block diagrams

% EAM modulator (absorption curve + transfer function)

% Table with IM/DD system parameters
\begin{table}[h]
    \centering
    \begin{tabular}{llc}
        \toprule
        \textbf{Component} & \textbf{Parameter Name} & \textbf{Value} \\
        \midrule
        \multirow{3}{*}{System} & Samples per symbol ($sps$) & 8 \\
        & Symbol rate ($R_s$) & 100 GBd \\
        & Constellation & PAM-4 \\
        & Bit rate (with 5.8\% FEC overhead) & 189 GBit/s \\
        \midrule
        \multirow{4}{*}{DAC} & DAC gain ($g_\text{dac}$) & - \\
        & 3dB cutoff frequency ($f_\text{3dB}$) & 45 GHz \\
        & Bias voltage ($V_\text{b}$) & - \\
        & Voltage peak-to-peak ($V_\text{pp}$) & - \\
        \midrule
        \multirow{3}{*}{Modulator} & Laser power ($P_\text{in}^{(\text{ideal})}$)& -13.0 dBm \\
        & Laser power ($P_\text{in}^{(\text{EAM})}$)& -6.0 dBm \\
        & Linewidth enhancement ($\alpha$) & 1.0 \\
        & Laser wavelength ($\lambda$) & 1270 nm \\
        \midrule
        \multirow{4}{*}{Fiber} & Dispersion slope ($S_0$) & 0.092 ps/(nm²·km) \\
        & Zero-dispersion wavelength ($\lambda_0$) & 1310 nm \\
        & Attenuation ($\alpha_\text{smf}$) & 0.2 dB/km \\
        & Dispersion parameter ($D$) & -15.43 ps/(nm·km) \\
        \midrule
        \multirow{6}{*}{Photodiode} & Boltzmann constant ($k$) & $1.38 \times 10^{-23}$ J/K \\
        & Temperature ($T$) & 293 K \\
        & Bandwidth ($B$) & 45 GHz \\
        & Impedance load ($Z$) & 50 $\Omega$ \\
        & Electron charge ($e_\text{c}$) & $1.6 \times 10^{-19}$ C \\
        & Responsivity ($R$) & 1 A/W \\
        & Dark current ($I_\text{d}$) & $1 \times 10^{-8}$ A \\
        \midrule
        \multirow{1}{*}{ADC} & 3dB cutoff frequency ($f_\text{3dB}$) & 45 GHz \\
        \bottomrule
    \end{tabular}
    \caption{Parameters used in the IM/DD system. Parameter values marked with '-' are either learned or varied during simulations.}
    \label{tab:imdd_system_parameters}
\end{table}

\subsection{Wavelength division multiplexing}
The minimization of the MSE, which is used for learning the pulse-shaper and receiver filter coefficients, does not put any constraints on the bandwidth of the learned filters directly. It is thereby expected that filters will occupy parts of the frequency spectrum outside the information bandwidth \cite{songModelBasedEndtoEndLearning2022} (see also Section~\ref{sec:results_awgn}). This will in turn make this framework much less desirable for wavelength division multiplexing (WDM) applications. 

\begin{figure}
    \centering
    \includegraphics[width=0.49\textwidth]{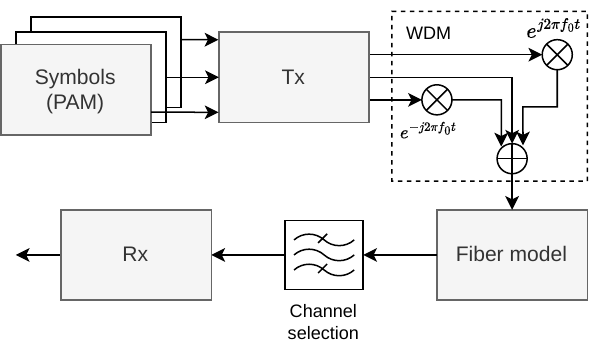}
    \caption{Wavelength division multiplexing (WDM) applied to both the AWGN and IM/DD systems during training and evaluation. The same pulse-shaping filter is applied to all channels in the Tx. We only decode the channel centered at 0 Hz.}\label{fig:wdm}
\end{figure}

To investigate this, we implemented a WDM scheme with three channels---one channel of interest, on which the pulse-shaper and receiver filter coefficients are learned and two interfering channels (see Figure~\ref{fig:wdm}). This scheme was employed both during training and evaluation to make the learned filters robust towards neighboring channels. Three independent symbol sequences are generated and processed by the transmitter. Note that the same, learned pulse-shaping filter, is applied to all three channels. The three independently modulated optical signals are then multiplexed with channel spacing $f_0$ after which the fiber model (either AWGN or IM/DD) is applied. After the fiber, we isolate the middle channel using a 5th order Gaussian frequency domain filter with a given bandwidth (see section~\ref{sec:results}). Receiver side processing is then applied, after which we calculate the SER (for the middle channel only).

% Generic WDM block

\section{Results}\label{sec:results}
We now investigate the performance of jointly optimizing the pulse-shaper and the receiver filter within the two simulation models described. To make a fair comparison, we consider the following four configuration variants:
\begin{description}[leftmargin=3\parindent,labelwidth=!]
    \item[\textbf{PS}] where the pulse-shaping filter is learned and the receiver filter is held fixed at RRC.
    \item[\textbf{RxF}] where the pulse-shaping filter is held fixed at RRC and the receiver filter is learned.
    \item[\textbf{PS \& RxF}] where both pulse-shaping filter and receiver filter are learned.
    \item[\textbf{RRC \& FFE}] where both pulse-shaping filter and receiver filter are held fixed at RRC and a linear feed-forward equaliser (FFE) is learned after the receiver filter but before downsampling.
\end{description}
During simulations, we vary the length of the filters but always keep the pulse-shaping filter and receiver filter at the same number of taps $N$. In the case where an equaliser is used (RRC \& FFE), we run the equaliser in the oversampled domain (i.e. samples per symbol is equal to the oversampling rate) and use $N$ (additional) taps. 

\subsection{Additive White Gaussian Noise Channel with Bandwidth Limitation}\label{sec:results_awgn}
\begin{figure*}
    \centering
    \includegraphics[width=\textwidth]{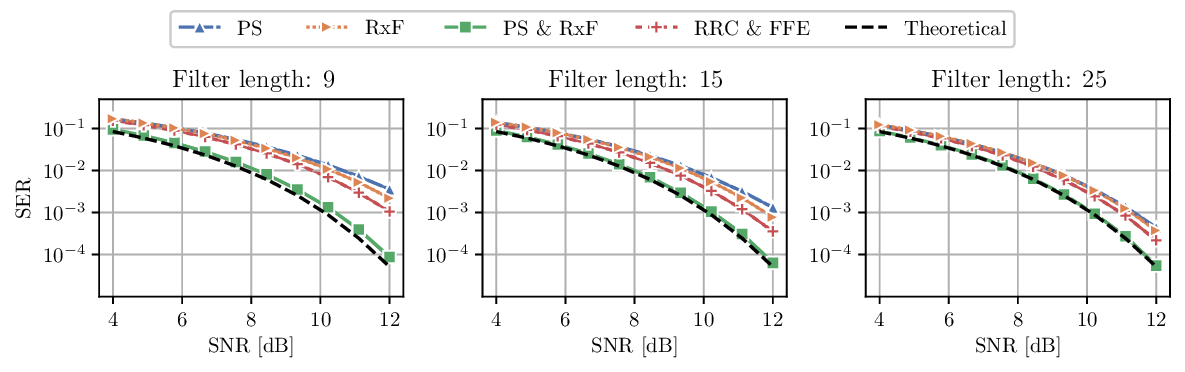}
    \caption{SER a function SNR for different filter lengths in the AWGN channel with bandwidth limitation. Filters were trained at SNR = 12 dB and evaluated in the range from 4 to 12 dB}
    \label{fig:awgn-ser-snr}
\end{figure*}

We compare the four learning methods on the AWGN channel described in section~\ref{sec:methods_awgn}. We vary the SNR and plot the SER for different filter lengths, which can be seen in Figure~\ref{fig:awgn-ser-snr}. Note that all filter optimization is carried out at an SNR of 12 dB. The choice of SNR does not influence the learned filters as the channel transfer function is linear. The theoretical curve is based on an AWGN channel without bandwidth limitation. The joint optimization framework (PS \& RF) comes extremely close to the theoretical limit even for very short filter lengths. The penalty to the theory almost vanishes with a filter length of $N = 25$. All other methods incur substantial penalties irrespective of the filter length. The baseline (RRC \& FFE) has a slight advantage over the other single-sided methods, which is attributed to the extra taps in the FFE. However, this advantage reduces as we increase the filter lengths.

\begin{figure*}
    \centering
    \subfloat[Pulse-shaping filter amplitude response]{\label{fig:awgn-sys-ps}\includegraphics[width=0.48\textwidth]{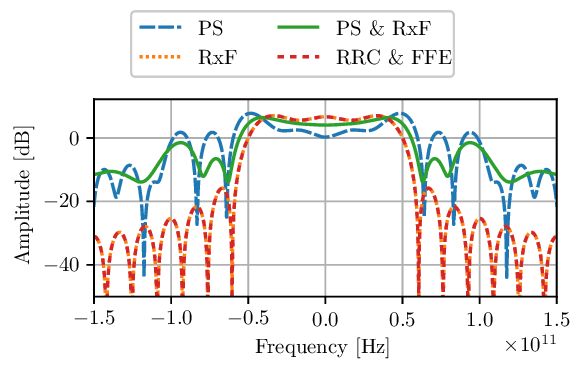}}
    \subfloat[Nyquist zero ISI condition (total system response).]{\label{fig:awgn-sys-zero-isi}\includegraphics[width=0.48\textwidth]{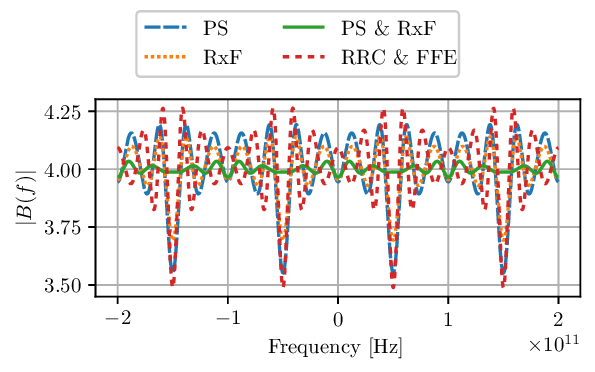}}
    \caption{AWGN: We show the amplitude spectrum of the pulse-shaper in~\ref{fig:awgn-sys-ps} for the different configurations. In~\ref{fig:awgn-sys-zero-isi}, we show the Nyquist zero ISI criterion. In this simulation, filter lengths were set to 25 for both pulse-shaper and receiver filter.}
    \label{fig:awgn-sys}
\end{figure*}

To illustrate what the different methods learn, we show the pulse-shaping amplitude response and the Nyquist zero ISI criterion~\cite{proakisDigitalCommunications2008} for the total system response (i.e. pulse-shaper, DAC, receiver filter and ADC) in Figure~\ref{fig:awgn-sys}. Given the total frequency response of the system, $H(f)$, then we are interested in analyzing the following term,
\begin{equation}
    B(f) = \sum_{m=-\infty}^\infty H\left(f + \frac{m}{T}\right).\label{eq:zero_isi}
\end{equation}
in which $T$ is the sampling interval. In an ISI-free system the quantity $B(f)$ is constant over all frequencies~\cite{proakisDigitalCommunications2008}. We see in Figure~\ref{fig:awgn-sys-zero-isi} that the joint optimization scheme achieves the flattest curve, which explains the near-optimal symbol-error performance in the AWGN channel (see Figure~\ref{fig:awgn-ser-snr}). Furthermore, we see in Figure~\ref{fig:awgn-sys-ps} that the methods that learn the pulse-shaper (PS and PS \& RxF) utilize a much wider range of frequencies compared to the RRC filter. This is due to the unconstrained nature of our cost function in the optimization problem~\cite{songModelBasedEndtoEndLearning2022}.

\subsection{Additive White Gaussian Noise Channel with Bandwidth Limitation and Wavelength Division Multiplexing}
\begin{figure*}
    \centering
    \subfloat[SER as a function SNR with channel spacing $f_0 = 150$ GHz.]{\includegraphics[width=\textwidth]{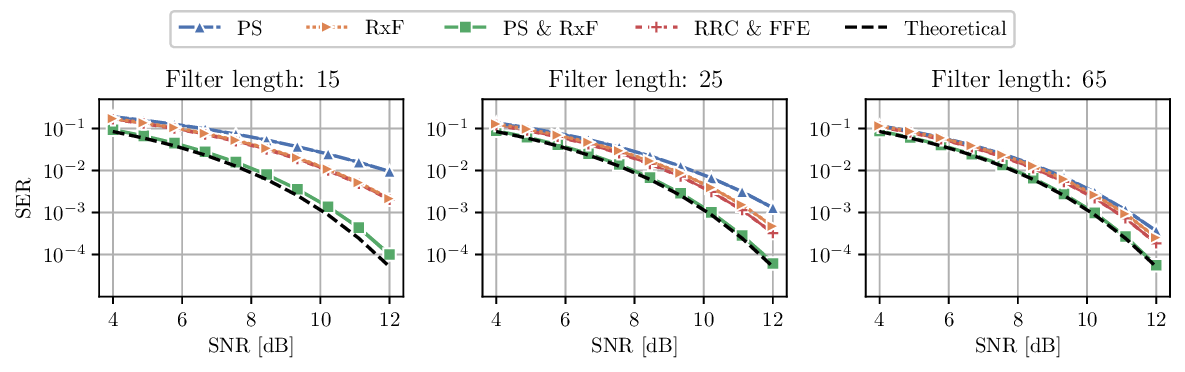}\label{fig:awgn-wdm-ser-snr}}\\
    \subfloat[SER as a function of channel spacing for fixed SNR of 12 dB]{\includegraphics[width=\textwidth]{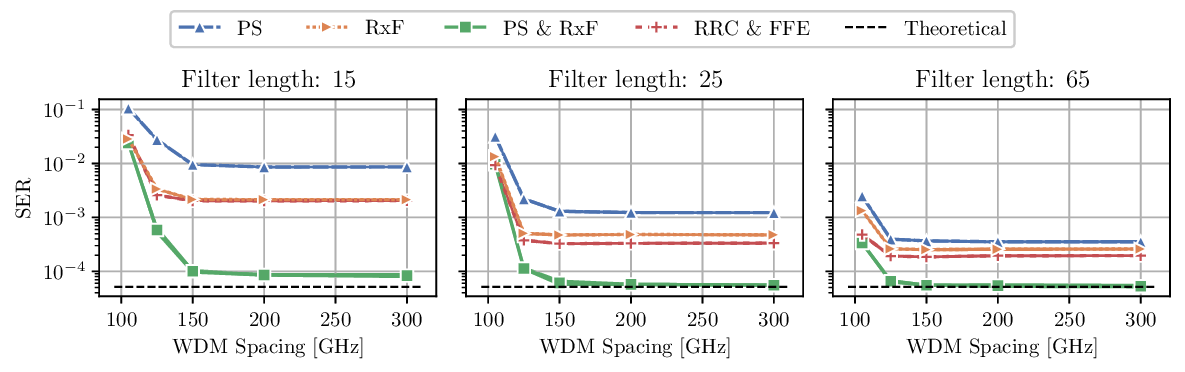}\label{fig:awgn-wdm-ser-chan-space}}
    \caption{Performance plots for the AWGN channel with bandwidth limitation and wavelength division multiplexing. In Figure~\ref{fig:awgn-wdm-ser-snr}, we have varied the SNR for a fixed channel spacing of $f_0 = 150$ GHz and in Figure~\ref{fig:awgn-wdm-ser-chan-space} we fixed the SNR to 12 dB and varied the channel spacing.}
    \label{fig:awgn-wdm}
\end{figure*}

We now study what happens if we include WDM in both training and evaluation and vary the channel spacing $f_0$. To accommodate for three channels, each with bandwidth of 100 GHz, we increase the oversampling rate from 4 to 8. We set the bandwidth of the Gaussian channel selection filter to be $f_\text{3dB} = 65$ GHz. We tried different values for $f_\text{3dB}$ between 55 and 65 GHz and the relative performance between methods did not significantly change. The performance of the systems can be seen in Figure~\ref{fig:awgn-wdm}. For a fixed channel spacing of 150 GHz (Figure~\ref{fig:awgn-wdm-ser-snr}), we see that the relative performance of the methods is similar to the previous section. Again, the join-optimization scheme is consistently better and can close the gap to the theoretical limit with a filter length of 65.
If we look at the performance when varying the channel spacing (Figure~\ref{fig:awgn-wdm-ser-chan-space}), we naturally see a degradation as the channels are spaced more closely together. The joint optimization PS \& RxF is still consistently better compared to the other configurations across the board. However, the benefit of optimizing both the PS and RxF filters jointly is only marginal in the low channel spacing regime ($<125$ GHz).

% AWGN WDM SER vs SNR for fixed f_0
% and
% AWGN WDM SER vs f_0

\subsection{Intensity Modulated and Direct Detected System}
% \input{imdd_system_parameters_table}
% IM/DD with WDM ideal - SER vs SNR
\begin{figure*}
    \centering
    \includegraphics[width=\textwidth]{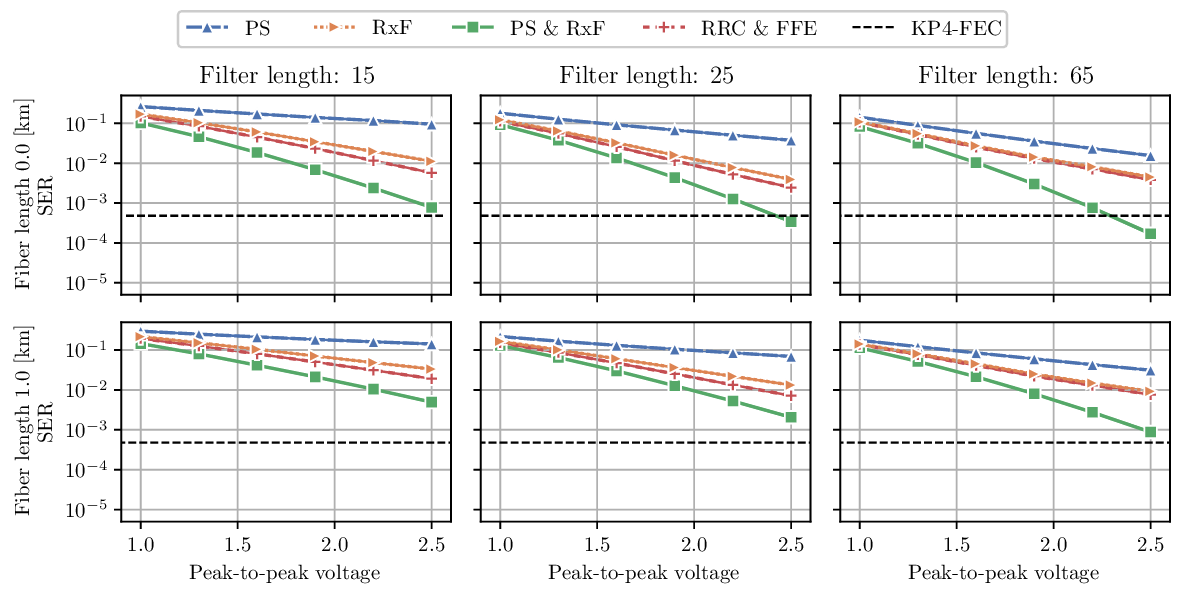}
    \caption{IM/DD system with an ideal linear modulator and channel spacing 200 GHz: SER vs $V_\text{pp}$ in DAC for different fiber and filter lengths.}
    \label{fig:imdd-ideal-ser-vpp}
\end{figure*}

% IM/DD with WDM ideal: SER vs channel spacing
\begin{figure*}
    \centering
    \includegraphics[width=\textwidth]{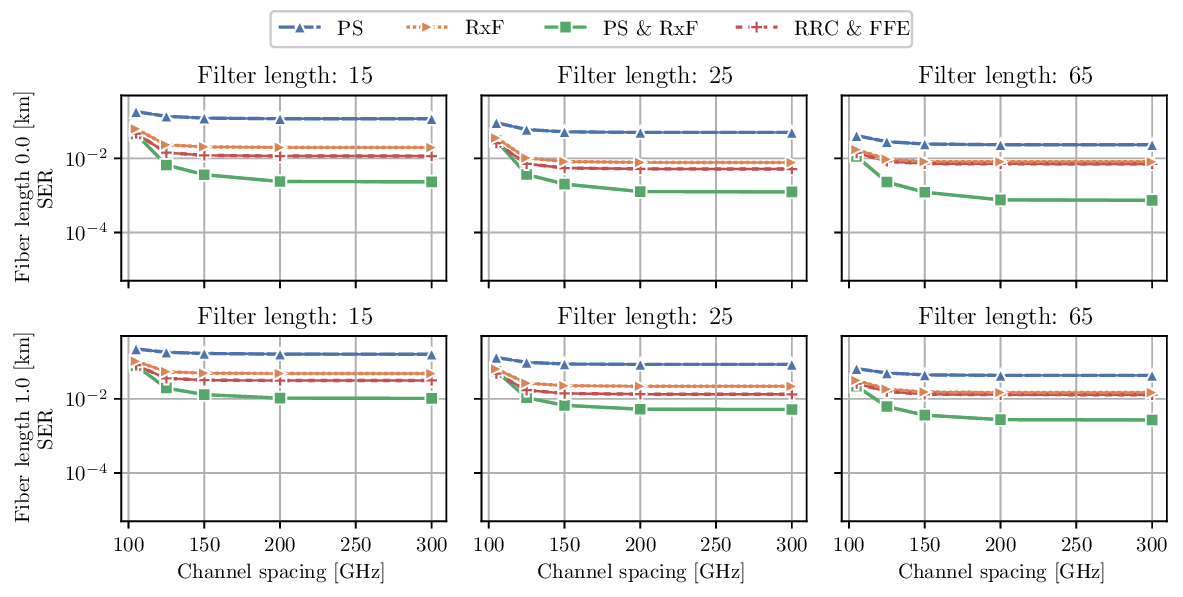}
    \caption{IM/DD system with ideal modulator: SER vs WDM channel spacing for different fiber and filter lengths with $V_\text{pp} = 2.2$ in DAC.}
    \label{fig:imdd-ideal-ser-chan-space}
\end{figure*}

We now turn to IM/DD system with WDM. During simulations we vary the peak-to-peak voltage, $V_\text{pp}$, in the DAC while keeping laser power constant. We test the system in a back-to-back (B2B) scenario and for a fiber length of 1.0 km. In all simulations, we learn the normalization constant $g_\text{dac}$ as part of the optimization such that all methods utilize the full voltage range. For the IM/DD system, we compare the performance of our methods to the pre-forward error correction threshold KP4-FEC, which we in symbol-error rate approximate by $\text{KP4}_\text{SER} \approx \log_2(4) \cdot 2.4\cdot 10^{-4}$.
The results of the numerical simulations with an ideal modulator can be seen in Figure~\ref{fig:imdd-ideal-ser-vpp} and \ref{fig:imdd-ideal-ser-chan-space}. For the ideal modulator we fix the bias in the DAC such that the voltages have range $[0, V_\text{pp}]$. We set the bandwidth of the Gaussian channel selection filter to be $f_\text{3dB} = 55$ GHz. Due to the ideal modulator, we observe a linear improvement in the SER as a function of $V_\text{pp}$ for all the methods, with the PS \& RxF method having the best performance. As in the AWGN simulation, when varying the channel spacing we need at least a channel spacing of 125 GHz to see a meaningful relative improvement for the joint optimization method compared to the others. We note that the receiver-side methods (RxF and RRC \& FFE) slightly \emph{increase} in SER for higher filter lengths. We attribute this to the clipping constraint in the DAC, which will penalize filters with higher peak-to-average power ratio (PAPR). Because we increase the filter length of the pulse-shaper and receiver filter jointly, this will create pulse-shaper filters with higher and higher PAPR (due to the very low RRC rolloff $\rho = 0.01$). The methods that optimize the pulse-shaper have the flexibility to learn filters with lower PAPR that avoid clipping and thus do not suffer an SER penalty at higher filter lengths.

% IM/DD with WDM and EAM - SER vs SNR
\begin{figure*}
    \centering
    \includegraphics[width=\textwidth]{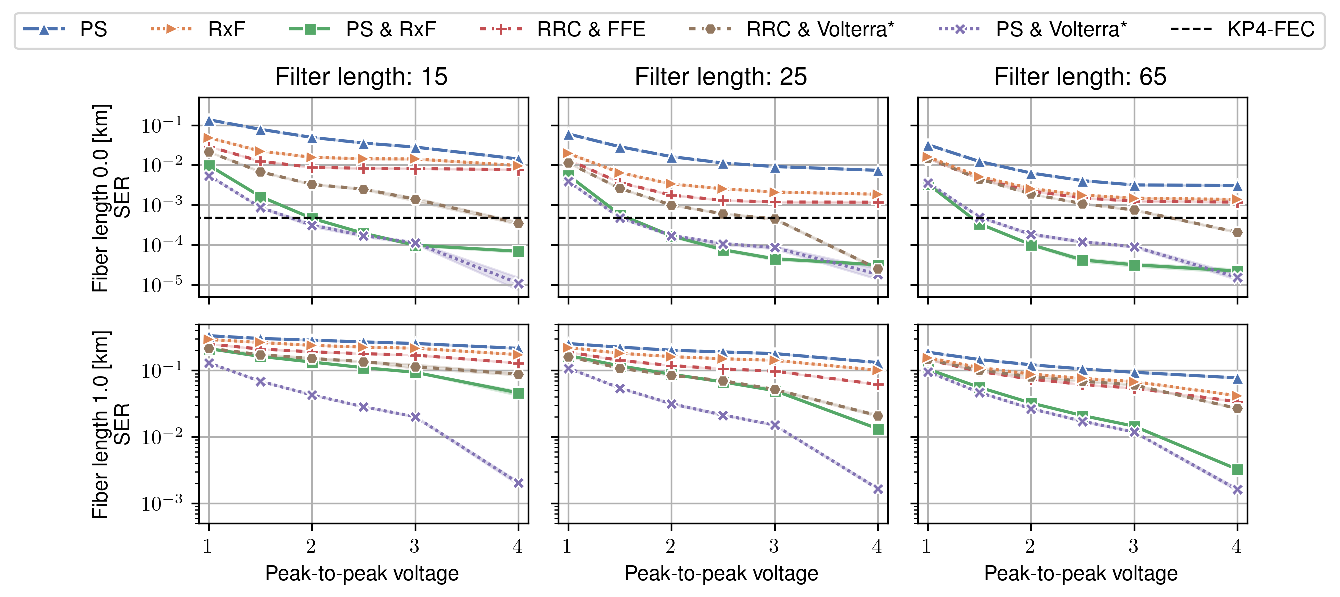}
    \caption{IM/DD system with EAM and channel spacing 200 GHz: SER vs $V_\text{pp}$ in DAC for different fiber and filter lengths. Note, the Volterra* methods has many more free parameters than the other methods and are only meant as a theoretical comparisons. The shaded area represents a 95\% confidence interval estimated over 5 random restarts of the simulation.}
    \label{fig:imdd-eam-ser-vpp}
\end{figure*}

% IM/DD with WDM EAM: SER vs channel spacing
\begin{figure*}
    \centering
    \includegraphics[width=\textwidth]{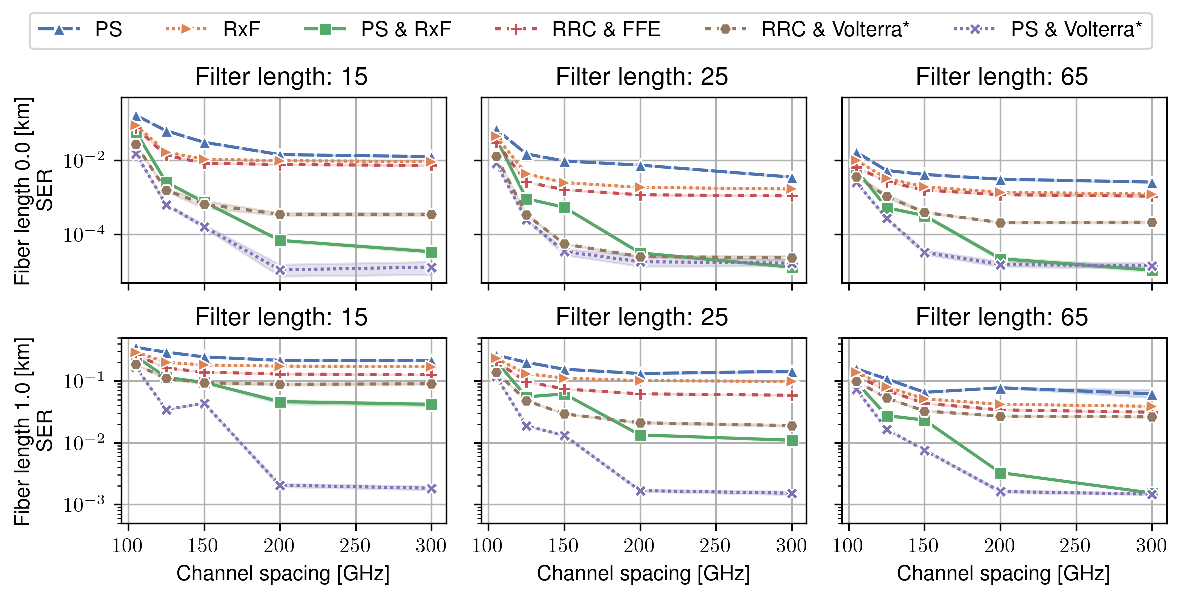}
    \caption{IM/DD system with EAM modulator: SER vs WDM channel spacing for different fiber and filter lengths with $V_\text{pp} = 4.0$ in DAC. Note, the Volterra* methods has many more free parameters than the other methods and are only meant as a theoretical comparisons. The shaded area represents a 95\% confidence interval estimated over 5 random restarts of the simulation.}
    \label{fig:imdd-eam-ser-chan-space}
\end{figure*}

\begin{figure*}
    \subfloat[Optimized pulse-shaper]{\includegraphics[width=0.49\textwidth]{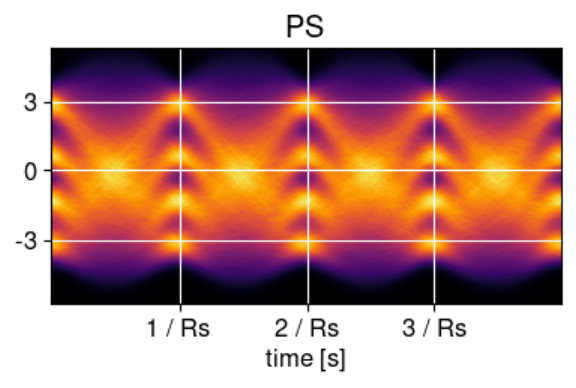}}
    \subfloat[Optimized receiver filter]{\includegraphics[width=0.49\textwidth]{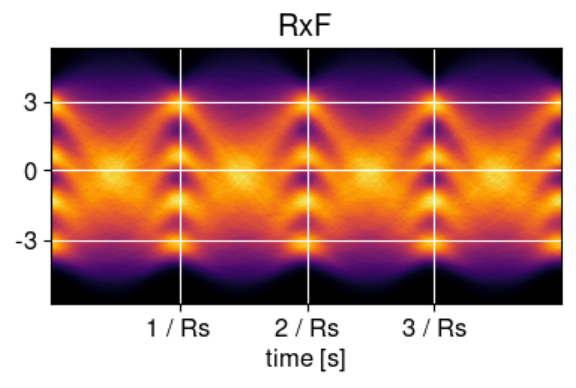}}\\
    \subfloat[Joint optimization of pulse-shaper and receiver filter]{\includegraphics[width=0.49\textwidth]{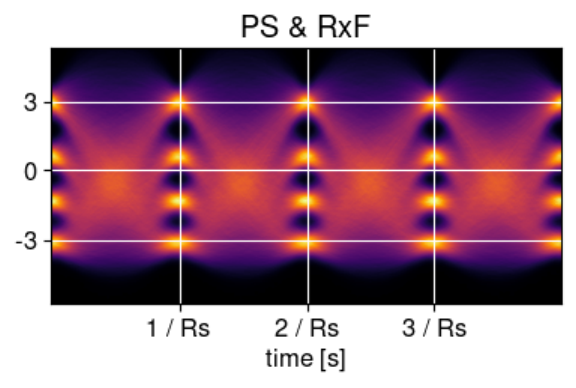}}
    \subfloat[RRC filters plus a linear FFE with 65 taps.]{\includegraphics[width=0.49\textwidth]{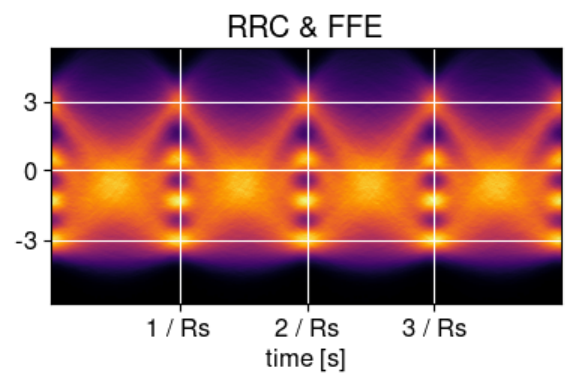}}
    \caption{Eyediagrams of the signal just before the downsampling in the IM/DD system. Fiber length was 1.0 km, $V_\text{pp} = 4.0$ and filter lengths of 65 (with an additional 65 equalizer taps in the FFE case).}
    \label{fig:imdd-eye}
\end{figure*}

% IM/DD with non-linear modulator
Finally, we investigate the learning framework with a non-linear modulator, namely the EAM (see input-output characteristic in Figure~\ref{fig:eam_transfer}). In all simulations with the EAM, in addition to the filters, we optimize $g_\text{dac}$ and the bias, $V_\text{b}$, in the DAC such that each method has the flexibility to utilize the modulator range in the best way possible. We initialized the bias in modulator to $V_\text{b} = -1.0$ for all simulations as we found that to yield consistently good results. Due to the non-linear modulator characteristic, we additionally compare the filter optimization methods with a second order Volterra equalizer. We set the number of taps in the first order filter to 101 and 45 taps in the second order, yielding 1136 free parameters. We use the non-linear equalizer in two configurations --- 1) together with a fixed RRC filter on the Tx side and a fixed RRC filter on the Rx side denoted RRC \& Volterra* and 2) together with a learnable pulse-shaper and fixed RRC filter on the Rx side denoted PS \& Volterra*. The two Volterra* methods can viewed as a best-case comparison, where we have not taken into account computational complexity constraints. The results of the numerical simulations can be seen in Figure~\ref{fig:imdd-eam-ser-vpp} and \ref{fig:imdd-eam-ser-chan-space}. In the B2B condition in Figure~\ref{fig:imdd-eam-ser-vpp}, due to the non-linear characteristic of the EAM, we see that the SER curve flattens as we increase the peak-to-peak voltage, except for the Volterra* methods that continue to improve. The relative performance between the methods is consistent with previous results, namely that the joint-learning framework outperforms the "single-side" frameworks.  One might expect that the Volterra equalizer alone (RRC \& Volterra*) should be enough to reach the optimum performance, with a clear advantage in the number of free parameters over the other standard filtering methods (PS, RxF and PS \& RxF). We attribute this to the ISI of the system stemming from bandwidth limitation being the limiting factor, and \emph{not} the non-linearity of the system. In a channel where the non-linearities are more dominant we would expect the Volterra alone to yield better results, but this was not the focus of the paper. If we additionally optimize the pulse-shaper, as represented by the PS \& Volterra method, we observe further gains in terms of performance compared to the linear approach (PS \& RxF). When we increase the fiber length to 1 km, we see a dramatic drop in performance and more linear behavior as a function of $V_\text{pp}$. We primarily attribute this to the phase noise introduced by the chirp in the modulator that significantly impacts the performance of all methods. Similar to the other simulations, the channel spacing in the WDM needs to be at least 125 GHz for the end-to-end learning framework to have an advantage over the other methods (see Figure~\ref{fig:imdd-eam-ser-chan-space}). However, we note that the for a fiber length of 1 km all methods struggle to reach good SER performance and the difference over channel spacing is relatively low (especially for short filters). As in the simulations with the ideal modulator, the "receiver-side only" methods (RxF, RRC \& FFE and RRC \& Volterra*) all incur a penalty when increasing the filter length from 25 to 65. As already discussed in the previous simulation, we attribute this to the clipping constraint in the DAC and the high PAPR of the RRC filter at the Tx. An example of eyediagrams after receiver filtering in the 1 km case can be seen in Figure~\ref{fig:imdd-eye}. The joint optimization (PS \& RxF) achieves the clearest eye-opening compared to the other methods.

% IM/DD with EAM: Eyediagrams

\subsubsection{Filter Robustness Evaluation with Non-linear Fiber Model}
Robustness of end-to-end frameworks will always be of interest as deviations from the simulation model will lead to performance degradation, unless specifically accounted for during training~\cite{karanovEndtoEndDeepLearning2018, jovanovicEndtoEndLearningConstellation2022}. To investigate the learned pulse-shaper and receiver filters' robustness and generalizability, we devised a small simulation in which the channel conditions are altered during evaluation.

\begin{figure*}
    \centering
    \includegraphics[width=\textwidth]{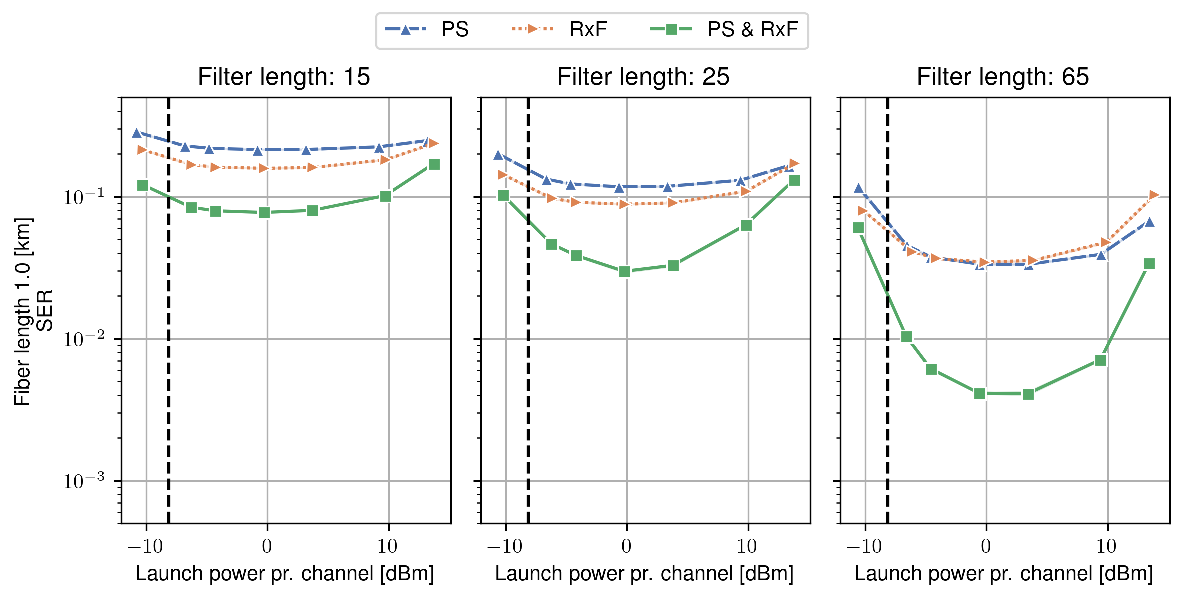}
    \caption{Robustness simulation: The plots show the evaluation performance of the IM/DD WDM (5 channel) systems with a non-linear fiber ($\gamma = 1.3$ [1/W/km]) as a function of laser input power ($P_{\text{in}}^{(\text{EAM})}$). The systems were trained with 3 channels on a standard SMF at $P_{\text{in}}^{(\text{EAM})} = -6.0$ dBm, which in launch power is indicated by the dashed black vertical line. During both training and evaluation the channel spacing was set to 150 GHz.}
    \label{fig:imdd-robustness}
\end{figure*}

The E2E systems under test, are trained on the IM/DD WDM system described above with a 1 km SMF with parameters described in Table~\ref{tab:imdd_system_parameters}. We used 3 WDM channels with a channel spacing of 150 GHz. During evaluation, we introduce a non-linear fiber with the split-step Fourier method (SSFM) with a non-linearity coefficient of $\gamma = 1.3$ [1/W/km]. The step size was set to $0.25$ km and we did not use any amplification. We introduce 2 additional interferer channels, bringing the total number of channels up to 5, and sweep the launch power that enters the fiber. The results of this simulation for PS, RxF and PS \& RxF can be seen in Figure~\ref{fig:imdd-robustness}.
Overall, all the methods follow the same trend, namely that the SER decreases until the non-linear distortion from the fiber becomes dominant and we start to see the SER increase.
The joint optimization method is across the board the best method, even in the highly non-linear regime, however, it also sees the biggest increase in SER compared relative to the optimum performance.

\section{Conclusion}\label{sec:conclusion}
We investigated the use of end-to-end learning for joint optimization of pulse-shaping and reciver-side FIR filters to mitigate ISI in bandwidth-limited communication channels. By implementing the channels in PyTorch, with the use of automatic differentiation we numerically investigated the performance gains in two types of systems; an AWGN channel with bandwidth-limitation and an IM/DD short-reach link with a non-linear modulator. We compared joint optimization to their respective single-sided counterpart, fixing one side to a RRC filter and learning the other side. We consistently found that joint optimization outperformed the other methods across both channels and SNRs. In the IM/DD system, we found support for short filter lengths ($N_\text{taps} = 15$) being able to respect the KP4 forward-error correction threshold which none of the other linear methods came close to. We hypothesize that joint optimizations' superiority comes from the additional degrees of freedom at the Tx, which results in the channel response, including hardware components, being more efficiently invertible.
This points toward joint-optimization methods being able to deliver considerable savings in terms of operating IM/DD links, at the (minor) cost of some calibration and optimization.

% Outlook
% Mention of experimential feasibility
Our framework does not directly translate to an experimental setup, as it requires differentiating through the channel. However, we believe that a combination of good characterization of electrical and optical components together with approximate optimization routines will be able to realize the reported performance gains.

\bibliography{bibtex/bib/JLT_2024}

\end{document}